\documentclass[12pt]{article}
\usepackage{amssymb,latexsym,amsmath,amsthm,bm}
\usepackage{authblk}

\usepackage[dvips]{color}
\usepackage{graphicx}
\usepackage[mathscr]{eucal}
\usepackage{booktabs}

\def\Ib{{\boldsymbol I}}

\def\Vb{{\boldsymbol V}}

\def\Xb{{\boldsymbol X}}
\def\Yb{{\boldsymbol Y}}

\def\thetab{{\boldsymbol      \theta}}

\def\xb{{\boldsymbol x}}

\def\Pib{{\boldsymbol         \Pi}}

\graphicspath{{eps_figs/}}

\def\appendix{\par \setcounter{section}{0} \def\thesection{APPENDIX:}}
\newdimen\jot \jot=2mm

\newtheorem{thm}{Theorem}[section]

\newtheorem{rmk}[thm]{Remark}

\def\vec{{\sf vec}}

\def\Ib{{\bf I}}

\def\Xb{{\bf X}}
\def\Yb{{\bf Y}}\def\Pib{{\bf \Pi}}

\def\diag{{\rm diag}}

\newcommand{\argmax}{\mathop{\rm argmax}}

\def\0{\textbf{0}}
\def\1{\textbf{1}}

\begin{document}

% Title of paper
\title{Matrix Variate Logistic Regression Model with Application to EEG Data}

\author{HUNG HUNG\footnote{To whom correspondence should be
addressed.}\\
\textit{Institute of Epidemiology and Preventive Medicine, National Taiwan University, Taipei, Taiwan}\\
{\small hhung@ntu.edu.tw}\\

\

CHEN-CHIEN WANG\\
\textit{Institute of Statistical Science, Academia Sinica, Taipei,
Taiwan}}
\date{}

\maketitle

% Add a footnote for the corresponding author if one has been
% identified in the author list

\begin{abstract}
{Logistic regression has been widely applied in the field of
biomedical research for a long time. In some applications,
covariates of interest have a natural structure, such as being a
matrix, at the time of collection. The rows and columns of the
covariate matrix then have certain physical meanings, and they must
contain useful information regarding the response. If we simply
stack the covariate matrix as a vector and fit the conventional
logistic regression model, relevant information can be lost, and the
problem of inefficiency will arise. Motivated from these reasons, we
propose in this paper the matrix variate logistic (MV-logistic)
regression model. Advantages of MV-logistic regression model include
the preservation of the inherent matrix structure of covariates and
the parsimony of parameters needed. In the EEG Database Data Set, we
successfully extract the structural effects of covariate matrix, and
a high classification accuracy is achieved.}\\

\end{abstract}

\section{Introduction}
\label{sec1}

Logistic regression has been widely applied in the field of
biomedical research for a long time. It aims to model the logit
transformation of conditional probability of an event as a linear
combination of covariates. After obtaining the data, maximum
likelihood estimates (MLE) can then be used to conduct subsequent
statistical analysis such as prediction and interpretation. When the
number of covariates excesses the available sample size as is
usually the case in modern research, the penalized logistic
regression (Lee and Silvapulle, 1988; Le Cessie and Van Houwelingen,
1992) is proposed to estimate parameters by maximizing the penalized
log-likelihood function, and to avoid the problems due to high
dimensionality. Recently, penalized logistic regression has been
applied to the filed of biomedical research, such as cancer
classification, risk factor selection, and gene interaction
detection, etc. See Zhu and Hastie (2004) and Park and Hastie (2008)
among others. We refer the readers to Muculloch and others (2008)
and Hastie and others (2009) for details and further extensions of
logistic regression.

In some applications, covariates of interest have a natural matrix
structure at the time of collection. For example, research interest
focuses on predicting the disease status of a patient based on
his/her image of certain organ. In this situation, we actually
obtain the data set of the form $\{(Y_i,X_i)\}_{i=1}^n$ which are
random copies of $(Y,X)$, where $Y$ is a binary random variable with
value 1 indicating diseased and 0 otherwise, and $X$ is a $p\times
q$ covariate matrix representing the corresponding image of the
$i$-th patient. Another example, the Electroencephalography (EEG)
Database Data Set which will be analyzed in this article later,
concerns the relationship between the genetic predisposition and
alcoholism. In this study, $Y=1$ and $Y=0$ represent alcoholic and
control groups, respectively, and the covariate $X$ is a $256 \times
64$ matrix, where its $(i,j)$-th element $X_{(i,j)}$ is the
measurement of voltage value at time point $i$ and channel of
electrode $j$.

With matrix covariate, we usually stack $X$ column by column as a
$pq$-vector, say $\vec(X)$, and subsequent statistical analysis
proceeds in the usual way. Specifically, conventional logistic
regression model admits that each $X_{(i,j)}$ possesses its own
effect $\xi_{ij}$ on the response $Y$. Motivated from the matrix
structure of $X$, it is natural to store $\xi_{ij}$'s into a
$p\times q$ parameter matrix $\xi$. Then, conventional logistic
regression model takes the formulation
\begin{eqnarray}
{\rm
logit}\{P(Y=1|X)\}&=&\gamma+\sum_{i,j}\xi_{ij}X_{(i,j)}\nonumber\\
&=&\gamma+\vec(\xi)^T\vec(X),\label{model.full}
\end{eqnarray}
where $\gamma$ is the intercept term. It is obvious that
model~(\ref{model.full}) does not consider the inherent matrix
structure of $X$, as the corresponding parameter matrix $\xi$ enters
model~(\ref{model.full}) only through $\vec(\xi)$ while its matrix
structure is ignored. As $X$ is a matrix at the time of collection,
it is reasonable for $\xi$ to possess certain structure. For
example, those $X_{(i,j)}$'s in a common region or in the same
column (row) of $X$ may have similar effects on the response.
Directly fitting (\ref{model.full}) without considering the
structural relationships among $\xi_{ij}$'s will cause the problem
of inefficiency. Moreover, by ignoring the inherent matrix
structure, one can hardly identify the row and column effects from
their joint effect matrix $\xi$, when the structural effects of $X$
are the interest of the study. Another problem of
model~(\ref{model.full}) is that the number of parameters usually
becomes extremely large in comparison with the sample size. For
instance, we have $1+64\times 256=16385$ parameters for the EEG
Database Data Set when fitting model~(\ref{model.full}), while the
available sample size is 122 only. As high dimensionality makes
statistical inference procedure unstable, it becomes urgent to seek
a more efficient method in dealing with matrix covariate. The main
theme of this research is thus to overcome the above mentioned
problems, via incorporating the hidden structural information of $X$
into statistical modeling.

The rest of this article is organized as follows. Section~2
introduces the matrix variate logistic regression model. Its
statistical meaning is also discussed. Section~3 deals with the
asymptotic properties of our estimators and the implementation
algorithm. The proposed method is further evaluated through two
simulation studies in Section~4 and the EEG Database Data Set in
Section~5. Section~6 illustrates the extension of matrix variate
logistic regression model to the case of multiple classes. The paper
is ended with conclusions in Section~7.

\section{Matrix Variate Logistic Regression Model}
\label{sec2}

\subsection{Model specification}

%We have seen in Section~1 that model~(\ref{model.full}) does not
%consider the inherent matrix structure of covariates and parameters.
To incorporate the structural information into modeling, as
motivated from the matrix structure of $X$, we propose the matrix
variate logistic (MV-logistic) regression model
\begin{eqnarray}\label{model}
{\rm logit}\{P(Y=1|X)\}=\gamma+\alpha^TX\beta,
\end{eqnarray}
where $\alpha=(\alpha_1,\cdots,\alpha_p)^T$ and
$\beta=(\beta_1,\cdots,\beta_q)^T$ are the row and column parameter
vectors, respectively, and $\gamma$ is the intercept term. As the
matrix structure of $X$ is preserved in model~(\ref{model}), these
parameters have their own physical meanings. In the EEG Database
Data Set, for instance, $\alpha$ is interpreted as the effect of
different time points, and $\beta$ as the effect of different
channels. By fitting MV-logistic regression model, we are able to
extract the column (row) information of $X$ which will provide
further insights into the relationship between $Y$ and $X$.
%as will be demonstrated in the
%analysis of the EEG Database Data Set.
Note that under model~(\ref{model}), $(\alpha, \beta)$ can only be
identified up to a scale, since $(c^{-1}\alpha, c\beta)$ will result
in the same model for any constant $c\neq 0$. For the sake of
identifiability, without loss of generality, we assume $\alpha_1=1$
in our derivation (see Remark~\ref{rmk.rho} for further discussion).
Denote the rest of parameters in $\alpha$ by $\alpha^*$, i.e.,
$\alpha=(1,\alpha^{*T})^T$, and
$\theta=(\gamma,\alpha^{*T},\beta^T)^T$ are the parameters of
interest. We thus have $p+q$ free parameters contained in $\theta$,
while it is $pq+1$ in the conventional logistic regression
model~(\ref{model.full}). One can see that a merit of
model~(\ref{model}) is the parsimony of parameters used.
%while it is only $p+q$ in MV-logistic
%regression model.
Thus, when model (\ref{model}) is correctly specified, an efficiency
gain in is expected.

Adoption of model~(\ref{model}) is equivalent to modeling the
covariate-specific odds ratio $R_{ij}$ of $X_{(i,j)}$ while keeping
the rest covariates fixed as
\begin{eqnarray}\label{or}
\log(R_{ij})=\alpha_i\beta_j~\Leftrightarrow~
R_{ij}=\{\exp(\beta_j)\}^{\alpha_i}.
\end{eqnarray}
Thus, the effect of $X_{(i,j)}$ depends on the product
$\alpha_i\beta_j$ instead of $\alpha_i$ or $\beta_j$ solely. A
positive $\alpha_i\beta_j$ implies $R_{ij}>1$. Note that since we
set $\alpha_1=1$, if $\beta_j>0$ ($<0$), then $\alpha_i>1$ ($<1$)
indicates $R_{ij}>R_{1j}$.
%that $X_{(i,j)}$ has a larger odds ratio in comparison
%with $X_{(1,j)}$.
Take the EEG Database Data Set to exemplify again, relation
(\ref{or}) implies that each channel has its own baseline odds ratio
$\exp(\beta_j)$. Depending on the time point of being measured, say
$i$, the odds ratio is further modified by taking a power of
$\alpha_i$.

\begin{rmk}\label{rmk.rho}
Although at the population level there is no difference for which
$\alpha_i$ is set to one, it is crucial to practical implementation.
If $\alpha_i$ is near zero, setting $\alpha_i=1$ will lead to
unstable results. Here we provide an easy guidance to select the
baseline effect. Let $\rho_{ij}$ be the sample correlation
coefficient between $X_{(i,j)}$ and $Y$. We then set $\alpha_i=1$ if
$i=\argmax_k\{\sum_{j=1}^q|\rho_{kj}|:k=1,\cdots,p\}$. The intuition
is to choose the one as the baseline which is the most likely to be
correlated with the response. We find in our numerical studies this
simple approach performs well.
\end{rmk}

\begin{rmk}\label{rmk.std}
Standardization of covariate will not affect the final result of
logistic regression model except a change of scale in parameter
estimates, since the standard deviation of each covariate can be
absorbed into the corresponding parameter. In MV-logistic regression
model, however, we have $pq$ covariates but only $p+q$ free
parameters and, hence, it is generally impossible to absorb those
standard deviations into fewer parameters. In summary,
standardization of covariates will result in a different MV-logistic
regression model. We will further discuss this issue in Section~5.
\end{rmk}

\subsection{Statistical meaning of MV-logistic regression model}

This subsection devotes to investigating the statistical meaning of
MV-logistic regression model. We will assume the validity of the
general model~(\ref{model.full}) with $(\gamma_0,\xi_0)$ being the
true value of $(\gamma,\xi)$. One will see that fitting MV-logistic
regression model actually aims to estimate \textit{the best rank-1
approximation} of the true parameter matrix $\xi_0$, in the sense of
minimum Kullback-Leibler divergence (KL-divergence), or
equivalently, maximum likelihood. This observation supports the
applicability of MV-logistic regression model in practice.

%under which to study how MV-logistic regression performs. As will be
%seen later that MV-logistic regression is conceptually similar to
%singular value decomposition (SVD), which aims to extract\textit{
%the best rank-1 approximation} of the true parameter matrix $\xi_0$.
%This fact then supports the applicability of MV-logistic regression
%in many applications.

First observe that, as
$\alpha^TX\beta=\vec(\alpha\beta^T)^T\vec(X)$, MV-logistic
regression model (\ref{model}) is equivalent to the conventional
model~(\ref{model.full}) with the constraint $\xi=\alpha\beta^T$.
Thus, MV-logistic regression utilizes the matrix structure of $\xi$
and approximates it by a rank-1 matrix $\alpha\beta^T$ in model
fitting. Secondly, it is known that maximizing the likelihood
function is equivalent to minimizing the KL-divergence (Bickel and
Doksum, 2001). Let $f(y|X;\gamma,\xi)$ be the conditional
distribution function of $Y$ given $X$ under
model~(\ref{model.full}). The KL-divergence between
$(\gamma_0,\xi_0)$ and any $(\gamma,\xi)$ is defined as
\begin{eqnarray}
KL(\gamma_0,\xi_0\|\gamma,\xi)=E_X\left\{\int \left(\log
\frac{f(y|X;\gamma_0,\xi_0)}{f(y|X;\gamma,\xi)}\right)f(y|X;\gamma_0,\xi_0)dy\right\},\label{kl.div}
\end{eqnarray}
where $E_X(\cdot)$ takes expectation with respect to $X$. Combining
the above two facts, at the population level, fitting MV-logistic
regression to estimate $(\gamma,\alpha,\beta)$ is equivalent to
searching the minimizer of the minimization problem
\begin{eqnarray}
\min_{\gamma,\alpha,\beta}~KL(\gamma_0,\xi_0\|\gamma,\alpha\beta^T).\label{kl.mv}
\end{eqnarray}
Let $(\alpha_0,\beta_0)$ be the minimizer of $(\alpha,\beta)$ in
(\ref{kl.mv}). MV-logistic regression then aims to search the matrix
$\alpha_0\beta_0^T$, which is called the best rank-1 approximation
of the true parameter matrix $\xi_0$. Notice that the optimality
between $\xi_0$ and $\alpha_0\beta_0^T$ here is not measured by the
usual Frobenius norm, but the KL-divergence in (\ref{kl.mv}) instead
(see Remark~\ref{rmk.svd} for more explications). Therefore, when
$\xi_0=\alpha_0\beta_0^T$, MV-logistic regression must be more
efficient than the conventional approach to estimating $\xi_0$ by
the usual MLE argument. Even if $\xi_0$ does not equal
$\alpha_0\beta_0^T$, we are still in favor of MV-logistic
regression, especially when the sample size is relatively small. In
fact, there is a trade-off between ``correctness of model
specification'' and ``efficiency of estimation''. With limited
sample size available, instead of estimating the full parameter
matrix $\xi_0$ which will suffer from the problem of unstable
estimation, MV-logistic regression model aims to more efficiently
estimate the best rank-1 approximation of $\xi_0$.

Clearly, the performance of MV-logistic regression model relies on
``how well the true parameter matrix $\xi_0$ can be approximated by
a rank-1 matrix $\alpha_0\beta_0^T$''. As demonstrated in Section~5
that MV-logistic regression outperforms the conventional approach in
the EEG Database Data Set, we believe this condition for $\xi_0$ is
not restrictive based on the following two reasons. First, it is
reasonable to assume that most covariates have effect sizes near
zero in modern biomedical research and, hence, $\xi_0$ is plausible
to be a low-rank matrix. The second reason is based on the
characteristic of the underlying study. In the EEG study, for
instance, measurements at the same time points (channels) would have
similar effects. These facts reflect the potential of $\xi_0$ to be
well explained by low rank approximation and, hence, the validity of
MV-logistic regression model. The robustness of MV-logistic
regression model will be further studied in Section~4.2.

\begin{rmk}\label{rmk.svd}
Given the matrix $\xi_0$, the weighted rank-1 approximation problem
of $\xi_0$ with the weight matrix $W$ (Lu and others, 1997; Manton
and others, 2003) is to find the minimizer $(\alpha_0,\beta_0)$ of
the minimization problem
$\min_{\alpha,\beta}\|\xi_0-\alpha\beta^T\|_W^2$,
%\begin{eqnarray}
%\min_{\alpha,\beta}\|\xi_0-\alpha\beta^T\|_W^2,\label{kl.norm2}
%\end{eqnarray}
where $\|\cdot\|_W^2=\vec(\cdot)^TW\vec(\cdot)$. The resulting
matrix $\alpha_0\beta_0^T$ is called the best rank-1 approximation
of $\xi_0$ in the sense of minimum $\|\cdot\|_W$ norm. When
$W=I_{pq}$, the $\|\cdot\|_W$ norm reduces to the Frobenius norm,
and $\alpha_0\beta_0^T$ can be obtained by singular value
decomposition (SVD) of $\xi_0$. Paralleling to this idea, fitting
MV-logistic regression model (with (\ref{kl.mv}) being the
estimation criterion) can be treated as finding the best rank-1
approximation of $\xi_0$ in the sense of minimum KL-divergence
(\ref{kl.div}).
\end{rmk}

%\begin{rmk}\label{rmk.svd2}
%When $f(y|X;\gamma,\xi)$ represents the pdf of normal with mean
%$\gamma+\vec(\xi)^T\vec(X)$ and a constant variance, we have shown
%(see Supplementary Material for derivation) that the minimizer
%$(\alpha_0,\beta_0)$ of $(\alpha,\beta)$ in
%$\min_{\gamma,\alpha,\beta}KL(\gamma_0,\xi_0\|\gamma,\alpha\beta^T)$
%equals the minimizer of
%\begin{eqnarray}
%%\min_{\gamma,\alpha,\beta}KL(\gamma_0,\xi_0\|\gamma,\alpha\beta^T)&=&
%%\min_{\alpha,\beta}KL(\gamma_0,\xi_0\|\gamma_0,\alpha\beta^T)\nonumber\\
%%&=&
%\min_{\alpha,\beta}\|\xi_0-\alpha\beta^T\|_{\Sigma_X}^2,\label{kl.norm}
%\end{eqnarray}
%where $\|\xi_0-\alpha\beta^T\|_{\Sigma_X}^2=
%\vec(\xi_0-\alpha\beta^T)^T\Sigma_X\vec(\xi_0-\alpha\beta^T)$ and
%$\Sigma_X=\cov\{\vec(X)\}$. The minimization problem (\ref{kl.norm})
%is the weighted low-rank approximation problem of $\xi_0$ with the
%weight matrix $\Sigma_X$ \citep{Lu1997, Manton2003}
%%(Lu, Pei, and Wang, 1997; Manton, Mahony, and Hua, 2003).
%Its minimizer $\alpha_0\beta_0^T$ is called the best rank-1
%approximation of $\xi_0$ under the norm $\|\cdot\|_{\Sigma_X}$. When
%$\Sigma_X=I_{pq}$, $\alpha_0\beta_0^T$ is the best rank-1
%approximation of $\xi_0$ in the usual sense of Frobenius norm, and
%can be obtained by SVD of $\xi_0$. These observations motivate
%MV-logistic regression to be a generalization of finding ``the best
%rank-1 approximation of $\xi_0$ in the sense of KL-divergence'' from
%normal regression model to logistic regression model.
%\end{rmk}

\section{Statistical Inference Procedure}
\label{sec3}

Some notations are defined here for ease of reference. For any
function $g$, $g^{(k)}$ represents the $k$-th derivative of $g$ with
respect to its argument. The parameters of interest are
$\theta=(\gamma,\alpha^{*T},\beta^T)^T$. Denote $P(Y_i=1|X_i)$ by
$\pi_i=\pi(\theta|X_i)=\exp(\gamma+\alpha^TX_i\beta)\{1+\exp(\gamma+\alpha^TX_i\beta)\}^{-1}$.
%\begin{eqnarray}
%\pi_i=\pi(\theta|X_i)=\frac{\exp(\gamma+\alpha^TX_i\beta)}{1+\exp(\gamma+\alpha^TX_i\beta)}.
%\end{eqnarray}
Define $\Pib(\theta)=(\pi_1,\cdots,\pi_n)^T$ and $\Vb(\theta)={\rm
diag}(v_1,\cdots,v_n)$, where $v_i=\pi_i(1-\pi_i)$ is the
conditional variance of $Y_i$ given $X_i$. Let
$\Yb=(Y_1,\cdots,Y_n)^T$ and
$\Xb(\theta)=[~\Xb_1(\theta),\cdots,\Xb_n(\theta)~]^T$ be an
$n\times pq$ matrix, where $\Xb_i(\theta)=(1, \beta^TX_i^TC,
\alpha^TX_i)^T$, $C=\partial
\alpha/\partial\alpha^*=[\0_{p-1},\Ib_{p-1}]^T$, $\0_a$ is the
$a$-vector of zeros, and $\Ib_a$ is the $a\times a$ identity matrix.
One can treat $\Xb_i(\theta)$ as the working covariates of the
$i$-th subject, which will be used in the development of our method.
Note that $C$ depends on which element of $\alpha$ is set to one. As
$\alpha_1=1$ throughout the paper, the notation $C$ is used for
simplicity.

\subsection{Estimation and Implementation}

The estimation of $\theta$ mainly relies on maximum likelihood
method. Given the data set $\{(Y_i,X_i)\}_{i=1}^n$, the
log-likelihood function of $\theta$ is derived to be
\begin{eqnarray}\label{likelihood}
\ell(\theta)=\sum_{i=1}^nY_i(\gamma+\alpha^TX_i\beta)-\log\{1+\exp(\gamma+\alpha^TX_i\beta)\},
%&=&\sum_{i=1}^nY_i\log(\pi_i)+(1-Y_i)\log(1-\pi_i)\nonumber\\
\end{eqnarray}
and $\theta$ can be estimated by the maximizer of $\ell(\theta)$.

In modern research of biostatistics, however, an important issue is
the number of covariates could be large in comparison with the
sample size. This will make the estimation procedure unstable, or
traditional methodologies may even fail. As mentioned in the
previous section, the number of free parameters in
model~(\ref{model}) is $p+q$, while it is $pq+1$ in the conventional
logistic regression model. MV-logistic regression then suffers less
severity from high dimensionality. However, this can not entirely
avoid the problem of instability. As we have seen in the EEG
Database Data Set, the covariate $X$ is a $256 \times 64$ matrix
(i.e., 320 parameters in MV-logistic regression model), while there
are only 122 observations. To overcome the difficulty of high
dimensionality, Le Cessie and Van Houwelingen (1992)
%Le Cessie and Van Houwelingen (1992)
proposed the penalized logistic regression method. This motivates us
to further consider the penalized MV-logistic regression. Let
$J(\theta)\geq 0$ be a twice continuously differentiable penalty
function of $\theta$, and $\lambda\geq 0$ be the regularization
parameter. We propose to estimate $\theta$ by
\begin{eqnarray}
\hat\theta_{\lambda}=\argmax_{\theta}\ell_{\lambda}(\theta),
\end{eqnarray}
where $\ell_{\lambda}(\theta)=\ell(\theta)-\lambda J(\theta)$ is the
penalized log likelihood. There are many choices of $J(\cdot)$
depending on different research purposes, wherein
$J(\theta)=\|\theta\|^2/2$ and
$J(\theta)=(\|\alpha^*\|^2+\|\beta\|^2)/2$ are the most widely
applied ones. The difference between them is whether to put the
penalty on the intercept term $\gamma$ or not. The regularization
parameter $\lambda$ should also be determined in practice. A
commonly used approach is to select $\lambda$ through maximizing the
cross-validated classification accuracy. Other selection criteria
can be found in Le Cessie and Van Houwelingen (1992).
%Le Cessie and Van Houwelingen (1992).

%\begin{eqnarray}
%\ell_{\lambda}(\theta)&=&\ell(\theta)-\lambda J(\theta),
%\end{eqnarray}

Interpretations of the elements of $\hat\theta_\lambda$ have been
introduced in Section~2.1. Let
$\hat\theta_{\lambda}=(\hat\gamma,\hat\alpha^{*T},\hat\beta^T)^T$
and $\hat\alpha=(1,\hat\alpha^{*T})^T$. The odds ratio $R_{ij}$ in
(\ref{or}) is estimated by $\exp(\hat\alpha_i\hat\beta_j)$. We would
also be interested in the success probability $\pi(\theta|x)$ for
any given $p\times q$ matrix $x$, which can be estimated by
$\pi(\hat\theta_{\lambda}|x)$. A discrimination rule is then to
classify a subject with matrix covariate $x$ to the ``diseased''
group if $\pi(\hat\theta_{\lambda}|x)>0.5$ and the ``non-diseased''
group otherwise. Detailed inference procedures about $\theta$ and
$\pi(\theta|x)$ will be discussed in the next subsection.

We close this subsection by introducing the iterative Newton method
to obtain $\hat\theta_{\lambda}$. The gradient of
$\ell_\lambda(\theta)$ (with respect to $\theta$) is calculated to
be
\begin{eqnarray}
\ell_\lambda^{(1)}(\theta)=\ell^{(1)}(\theta)-\lambda
J^{(1)}(\theta), \label{grad}
\end{eqnarray}
where $\ell^{(1)}(\theta)=\Xb(\theta)^T\{\Yb-\Pib(\theta)\}$.
Moreover, the Hessian matrix of $\ell_\lambda(\theta)$ is derived to
be
\begin{eqnarray}
\ell_\lambda^{(2)}(\theta)=-H_\lambda(\theta)+\left[\begin{array}{ccc}0&0&0\\
0&0&\sum_{i=1}^nC^TX_i(Y_i-\pi_i)\\0&\sum_{i=1}^nX_i^TC(Y_i-\pi_i)&0\end{array}\right],
\label{hess0}
\end{eqnarray}
where
\begin{eqnarray}\label{hess}
H_\lambda(\theta)=\Xb(\theta)^T\Vb(\theta)\Xb(\theta)+\lambda
J^{(2)}(\theta).
\end{eqnarray}
As suggested by Green (1984), we will ignore the last term of
(\ref{hess0}) since its expectation is zero. Finally,
$\hat\theta_\lambda$ can be obtained through iterating
\begin{eqnarray}
\theta_{(k+1)}=\theta_{(k)}+\left\{H_\lambda(\theta_{(k)})\right\}^{-1}
\ell_\lambda^{(1)}(\theta_{(k)}),~k=0,1,2,\cdots,\label{newton}
\end{eqnarray}
until there is no significant difference between $\theta_{(k+1)}$
and $\theta_{(k)}$, and outputs
$\hat\theta_{\lambda}=\theta_{(k+1)}$. A zero initial $\theta_{(0)}$
is suggested and performs well in our numerical studies.

%\begin{rmk}
%We suggest using a larger penalty (e.g., $10\lambda$) at the first
%iteration to obtain an initial parameter value for the subsequent
%iterations. It is found in our numerical studies that this procedure
%fastens the convergence of iterative Newton method.
%\end{rmk}

\subsection{Asymptotic Properties}

Asymptotic properties of $\hat\theta_\lambda$ can be derived through
usual arguments of MLE. The result is summarized in
Theorem~\ref{normality} below.

\begin{thm}\label{normality}
Assume the validity of model~(\ref{model}) and the regularity
conditions of the likelihood function. Assume also the information
matrix $I(\theta)=E\{\Xb_i(\theta)v_i\Xb_i(\theta)^T\}$ is
nonsingular. Then, for any fixed $\lambda$,
$\sqrt{n}(\hat\theta_\lambda-\theta)$ converges weakly to
$N\left(0,\Sigma(\theta)\right)$, where
$\Sigma(\theta)=\{I(\theta)\}^{-1}$.
\end{thm}

From Theorem~\ref{normality}, $\hat\theta_\lambda$ is shown to be a
consistent estimator of $\theta$. It also enables us to construct a
confidence region of $\theta$, provided we have an estimate of
$\Sigma(\theta)$. This can be done by the usual empirical estimator
with the unknown $\theta$ being replaced by $\hat\theta_\lambda$. In
particular, define
\begin{eqnarray}
\hat\Sigma(\theta)=\left(\frac{1}{n}H_\lambda(\theta)\right)^{-1}\left(\frac{1}{n}\Xb(\theta)^T\Vb(\theta)\Xb(\theta)\right)
\left(\frac{1}{n}H_\lambda(\theta)\right)^{-1}.\label{asyp.cov}
\end{eqnarray}
We propose to estimate the asymptotic covariance matrix
$\Sigma(\theta)$ by $\hat\Sigma(\hat\theta_\lambda)$. For any
$0<a<1$, an approximate $100(1-a)\%$ confidence interval of
$\theta_i$, the $i$-th element of $\theta$, is constructed to be
\begin{eqnarray}\label{ci}
\left(\hat\theta_{\lambda,i}-
z_{\frac{a}{2}}\frac{[\hat\Sigma(\hat\theta_\lambda)]_{i}}{\sqrt{n}},~\hat\theta_{\lambda,i}+
z_{\frac{a}{2}}\frac{[\hat\Sigma(\hat\theta_\lambda)]_{i}}{\sqrt{n}}\right),
\end{eqnarray}
where $z_{a/2}$ is the $1-a/2$ quantile of standard normal and
$[\hat\Sigma(\hat\theta_\lambda)]_{i}$ denotes the $i$-th diagonal
element of $\hat\Sigma(\hat\theta_\lambda)$. As to making inference
about $\pi(\theta|x)$ for any $p\times q$ matrix $x$, by applying
delta method and the result of Theorem~\ref{normality}, we have
\begin{eqnarray}
\sqrt{n}\left\{\log\left(\frac{\pi(\hat\theta_{\lambda}|x)}{1-\pi(\hat\theta_{\lambda}|x)}\right)
-\log\left(\frac{\pi(\theta|x)}{1-\pi(\theta|x)}\right)\right\}\stackrel{d}{\rightarrow}
N\left(0,\sigma_{\pi}^2(\theta|x)\right),
\end{eqnarray}
where
$\sigma_{\pi}^2(\theta|x)=\xb(\theta)^T\Sigma(\theta)\xb(\theta)$
and $\xb(\theta)=(1,\beta^Tx^TC,\alpha^Tx)^T$. The asymptotic
variance $\sigma_{\pi}^2(\theta|x)$ can be estimated by
$\hat\sigma_{\pi}^2(\hat\theta_{\lambda}|x)$, where
$\hat\sigma_{\pi}^2(\theta|x)=
\xb(\theta)^T\hat\Sigma(\theta)\xb(\theta)$. An approximate
$100(1-a)\%$ confidence interval of $\pi(\theta|x)$ is then
constructed to be
\begin{eqnarray}\label{ci.pi}
\left(\frac{\exp(\hat\gamma+\hat\alpha^T
x\hat\beta-z_{\frac{a}{2}}\frac{\hat\sigma_{\pi}(\hat\theta_{\lambda}|x)}{\sqrt{n}})}{1+\exp(\hat\gamma+\hat\alpha^T
x\hat\beta-z_{\frac{a}{2}}\frac{\hat\sigma_{\pi}(\hat\theta_{\lambda}|x)}{\sqrt{n}})}~,~\frac{\exp(\hat\gamma+\hat\alpha^T
x\hat\beta+z_{\frac{a}{2}}\frac{\hat\sigma_{\pi}(\hat\theta_{\lambda}|x)}{\sqrt{n}})}{1+\exp(\hat\gamma+\hat\alpha^T
x\hat\beta+z_{\frac{a}{2}}\frac{\hat\sigma_{\pi}(\hat\theta_{\lambda}|x)}{\sqrt{n}})}\right)
\end{eqnarray}
which is guaranteed to be a subinterval of $[0,1]$.

%\begin{rmk}
%Theorem~\ref{normality} states the asymptotic result and we must
%have a bias term $\lambda_0b(\theta)$ in practical implementation.
%It is natural to correct the resulting estimator
%$\hat\theta_\lambda$ by subtracting an estimate of $b(\theta)$. We
%propose to estimate $b(\theta)$ by $\hat b(\hat\theta_\lambda)$,
%where
%\begin{eqnarray}
%\hat
%b(\theta)=\left(\frac{1}{n}H(\theta)\right)^{-1}J^{(1)}(\theta).
%\end{eqnarray}
%The bias-corrected estimator of $\theta$ is then proposed to be
%\begin{eqnarray}
%\hat\theta_{\lambda,\text{BC}}=\hat\theta_\lambda-\frac{\lambda}{n}\hat
%b(\hat\theta_\lambda).
%\end{eqnarray}
%\end{rmk}

\section{Simulation Studies}
\label{sec4}

The proposed method is evaluated through simulation under two
different settings. Let the parameters be set as $\gamma=1$,
$\alpha=(1,0.5,-0.5\1_{p-2}^T)^T$, and $\beta=(1,
0.5,1,-\1_{q-3}^T)^T$, where $\1_a$ is the $a$-vector of ones, and
the penalty function $J(\theta)=(\|\alpha^*\|^2+\|\beta\|^2)/2$ is
considered. The tuning parameters for different logistic regression
models are determined from an independent simulation by maximizing
the classification accuracy. Simulation results are reported with
1000 replicates.

\subsection{Simulation under MV-logistic regression model}

The first simulation study evaluates the proposed method under
model~(\ref{model}) with $(p,q)=(12,10)$. We first generate $X$ such
that $\vec(X)$ follows a $pq$-variate normal distribution with mean
zero and covariance matrix $\Ib_{pq}$. Conditional on $X$, $Y$ is
generated from model~(\ref{model}) with the specified $\theta$. The
averages of $\hat\theta_{\lambda}$ and standard errors from the
diagonal elements of $\hat\Sigma(\hat\theta_{\lambda})$ in
(\ref{asyp.cov}) are provided in Table~\ref{sim1} for $n=150,300$.
For the case of small sample size $n=150$, biases for
$\hat\theta_\lambda$ are detected. The biases are mainly due to the
penalty term $\lambda J(\theta)$, as is the case of ridge
regression. In fact, from the proof of Theorem~\ref{normality}, the
bias term is derived to be $\lambda
n^{-1}\{I(\theta)J^{(1)}(\theta)\}$ which is of order $n^{-1}$. The
biases, however, are all relatively small in comparison with the the
corresponding standard deviations of $\hat\theta_\lambda$, and
decrease as the sample size increases. Moreover, all the standard
deviations are well estimated by the diagonal elements of
$\hat\Sigma(\hat\theta_{\lambda})$. These observations validate
Theorem~\ref{normality} and the proposed empirical variance
estimator.

As
$\hat\gamma+\hat\alpha^TX\hat\beta=\hat\gamma+(\hat\beta^T\otimes\hat\alpha^T)\vec(X)$
is the critical component used in prediction, we also report the
averages of similarities $u^Tv/(\|u\|\|v\|)$ with
$u=(\gamma,\beta^T\otimes\alpha^T)^T$ and
$v=(\hat\gamma,\hat\beta^T\otimes\hat\alpha^T)^T$ in the last row of
Table~~\ref{sim1}. Although biases arise especially for the case of
$n=150$, the similarities are not affected and have values very
close to one in both cases. This means MV-logistic regression would
have good performance in classification, since it is the direction
of $(\gamma,\beta^T\otimes\alpha^T)^T$ that is relevant to
classification. To demonstrate this, in each simulation replicate we
also generate another independent data set, and calculate the
classification accuracy by using
$\hat\gamma+\hat\alpha^TX\hat\beta$. For comparison, we also fit the
conventional logistic regression model~(\ref{model.full}) to obtain
estimates of $(\gamma,\xi)$ as if we ignore the relationship
$\xi=\alpha\beta^T$, and denote it by $(\tilde\gamma,\tilde\xi)$.
The classification accuracy obtained from
$\tilde\gamma+\vec(\tilde\xi)^T\vec(X)$ is also calculated. The
averaged classification accuracies and the winning proportions (the
proportion of MV-logistic regression with higher classification
accuracy over 1000 replicates) under $n=150$ are placed in
Table~\ref{sim3} (the row indexed by $\sigma=0$). It is detected
that MV-logistic regression produces more accurate results, and the
superiority of MV-logistic regression becomes more obvious for
larger $(p,q)$ values.

\subsection{Simulation violating MV-logistic regression model}

In the previous numerical study, MV-logistic regression outperforms
conventional logistic regression when data is generated from
model~(\ref{model}). It is our purpose here to evaluate the
performance of MV-logistic regression, when the underlying
distribution departures from model~(\ref{model}). The same setting
in Section~4.1 is used, except for each simulation replicate, $Y$ is
generated from the conventional logistic regression
model~(\ref{model.full}) with $\gamma=1$ and
$\xi=\alpha\beta^T+\delta$, where each element of the $p\times q$
matrix $\delta$ is also randomly drawn from a normal distribution
with mean zero and variance $\sigma^2$. With an extra term $\delta$,
model~(\ref{model}) is violated, and the magnitude of violation is
controlled by $\sigma$. Both models (\ref{model.full}) and
(\ref{model}) are fitted to obtain the estimates
$(\tilde\gamma,\tilde\xi)$ and $(\hat\gamma,\hat\xi)$ with
$\hat\xi=\hat\alpha\hat\beta^T$, respectively. We compare the
classification accuracies of the predictors
$\hat\gamma+\hat\alpha^TX\hat\beta$ and
$\tilde\gamma+\vec(\tilde\xi)^T\vec(X)$ applied on another
independent data set, under different combinations of $(p,q)$ and
$\sigma=0.1,0.3,0.5$. Note that $\sigma=0$ means MV-logistic
regression is the true model. To see how these $\sigma$ values
affect the deviation from MV-logistic regression model, we also
report the corresponding explained proportions $\rho_\sigma$ defined
below. Let $k=p\wedge q$ and $s_1>\cdots>s_k>0$ be the $k$ non-zero
singular values of each generated $\xi$. Then $\rho_\sigma$ is the
average (over 1000 replicates) of $s_1/(\sum_{\ell=1}^ks_\ell)$.
Here $0<\rho_\sigma<1$ measures how well $\xi$ can be explained by
its best rank-1 approximation. Small value of $\rho_\sigma$
indicates severe violation of MV-logistic regression model.

The analysis results with $n=150$ are placed in Table~\ref{sim3}.
For every $\sigma$, conventional logistic regression is the correct
model, and its classification accuracies are irrelevant to $\sigma$
but will decay rapidly as $pq$ increases. On the other hand,
MV-logistic regression is less affected by $(p,q)$ than conventional
approach as it involves only $p+q$ parameters. The cost for its
parsimony of parameters is that, for any fixed $(p,q)$, its
performance becomes worse for larger $\sigma$. Overall, for moderate
deviations $\sigma\leq0.3$, MV-logistic regression outperforms the
conventional approach for every combination of $(p,q)$. For
$\sigma=0.5$, MV-logistic regression has lower classification
accuracy at $(p,q)=(12,10)$. We note that in this case,
$\rho_\sigma=33\%$ implies a large deviation from MV-logistic
regression model, while conventional logistic regression is expected
to have better performance as the number of parameters is
$121<n=150$. For larger $(p,q)$ values, however, MV-logistic
regression will be the winner even in the most extreme case of
$(p,q)=(20,30)$, where the explained proportion $\rho_\sigma$ is
$22\%$ only.
%Note that $\rho_\sigma$ is decreasing in $(p,q)$ for
%any fixed $\sigma$, which means the magnitude of model deviation
%becomes more severe for larger $(p,q)$.
It indicates that when $p$ and $q$ are large, the gain in efficiency
(from fitting MV-logistic regression model) more easily exceeds the
loss due to model misspecification. These observations show that
MV-logistic regression model has certain robustness against the
violation of model specification, especially for large number of
covariates.

\section{The EEG Database Data Set}
\label{sec5}

In this section, we analyze the EEG Database Data Set to demonstrate
the usefulness of MV-logistic regression. The data set consists of
122 subjects, wherein 77 of them belong to the group of alcoholism
$(Y_i=1)$ and the rest 45 subjects are in the control group
$(Y_i=0)$. Each subject completed a total of 120 trials under three
different conditions (single stimulus, two matched stimuli, and two
unmatched stimuli). In each trial, measurements from 64 electrodes
placed on subject's scalp at 256 time points are collected, which
results in a $256\times 64$ covariate matrix. It is interested to
distinguish two types of subjects based on the collected matrix
covariates. The data set can be downloaded from the web site of UCI
Machine Learning Repository
(\textit{http://archive.ics.uci.edu/ml/datasets/EEG+Database}).

\begin{figure}[!b]
\centering\includegraphics[height=12cm]{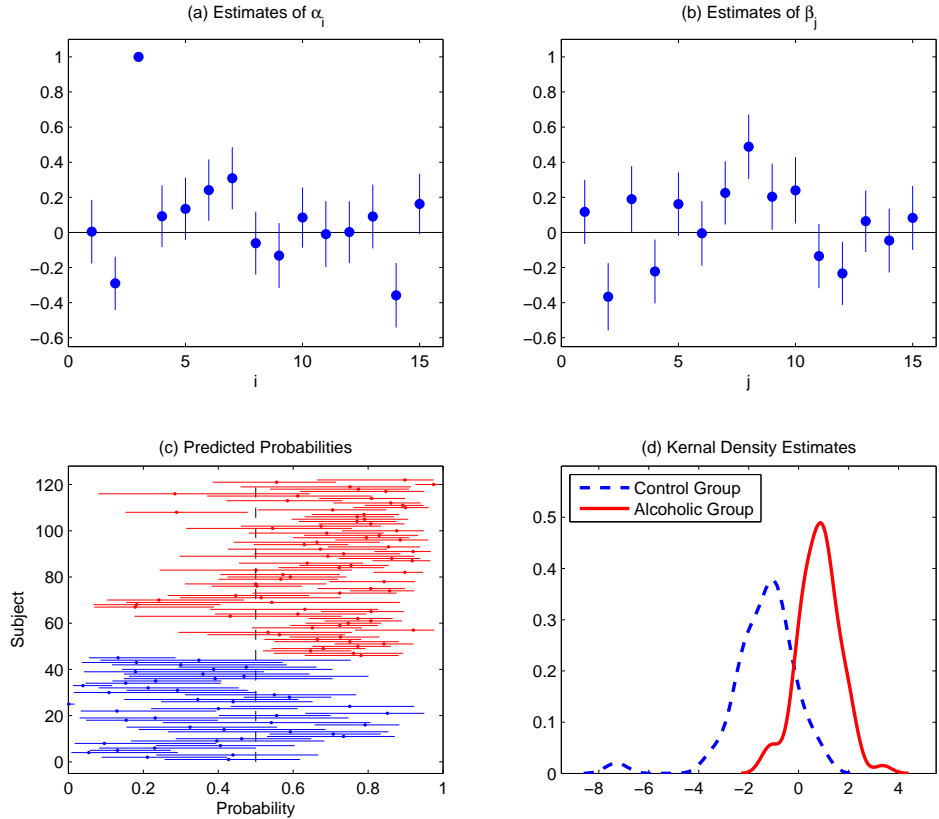}
\caption{Analysis results of the EEG Database Data Set with
$(p_0,q_0)=(15,15)$. (a)-(b) Estimates of $\alpha$ and $\beta$ (the
circles) with the $95\%$ confidence intervals (the vertical bars).
As we set $\alpha_3=1$, no confidence interval is provided for
$\alpha_3$. (c) Estimates of $\pi(\theta|X_i)$ (the symbol $*$) with
$95\%$ confidence intervals (the horizontal lines). Subjects 1--45
and 46--122 belong to the control and alcoholic groups,
respectively. The vertical dash line indicates a probability of
value 0.5. (d)~Kernel density estimates of
$\hat\alpha^TX_i\hat\beta$ for control and alcoholic groups.}
\label{eeg_estimates}
\end{figure}

The EEG Database Data Set was recently analyzed by Li and others
(2010)
%Li, Kim, and Altman (2010),
whose main purpose focused on dimension reduction. Here we adopt a
similar strategy for data preprocessing. In particular, we consider
partial data set of single-stimulus experimenters only, and the
averaged matrix covariates over different trials of the same subject
(denoted by $X_i^*$) will be considered in our analysis. This data
setting is same with the one used in Li and others (2010). Note that
with $256\times 64$ covariate matrix, we have 320 free parameters
which is still an excess of the sample size 122. Before fitting
MV-logistic regression model, the generalized low rank
approximations (GLRAM) of Ye (2005) is performed to reduce the
dimensionality of $X_i^*$ first. GLRAM is an extension of principal
component analysis to matrix objects, which aims to find orthogonal
bases $A\in \mathbb{R}^{p\times p_0}$ and $B\in\mathbb{R}^{q\times
q_0}$ with $p_0<p$ and $q_0<q$ such that $X_i^*$ is well explained
by the lower dimensional transformation $A^TX_i^*B$.
%Hung, Wu, Tu, and Huang (2011) have developed
%statistical justification of GLRAM.
Detailed analysis procedure is listed below.
\begin{itemize}
\item[1.]
Apply GLRAM to find $A$ and $B$ under $(p_0,q_0)=(15,15)$. Define
$\hat X_i^*=A^TX_i^*B$.
%such that $\sum_{i=1}^{122} \|(X_i^*-\bar X^*) -
%AA^T (X_i^*-\bar X^*) BB^T\|_F^2$ is minimized, where $\|\cdot\|_F$
%denotes the Frobenius norm of a matrix and $\bar X^*$ is the mean
%matrix of $X_i^*$.

\item[2.]
Standardize each element of $\hat X_i^*$ to obtain the covariate
matrix $X_i$.

\item[3.]
Fit MV-logistic regression with $\{(Y_i,X_i)\}_{i=1}^{122}$. We
apply the rule suggested in Remark~\ref{rmk.rho} to set $\alpha_3=1$
and denote the rest $\alpha_i$'s by $\alpha^*$.
\end{itemize}
To estimate $\theta$ in Step~3, we adopt the penalty function
$J(\theta)=\|\theta\|^2/2$. The penalty $\lambda=24$ is chosen so
that the leave-one-out classification accuracy is maximized. The
resulting estimates of $\alpha$ and $\beta$ are provided in
Figure~\ref{eeg_estimates}~(a)-(b) with the corresponding $95\%$
confidence intervals constructed from (\ref{ci}). As many estimates
of $\alpha_i$'s and $\beta_j$'s are significantly different from
zero, channels and measurement times surely play important roles in
distinguishing alcoholic and control groups. Observe that all the
estimates of $\alpha^*$ are smaller than 1. Thus, for those channels
with $\hat\beta_j>0$, the effects of measurement times are all
smaller than the third time point. In other words, most channels
achieve the largest odds ratio at the third time point. For the rest
channels with $\hat\beta_j<0$, the largest effect happens at the
$14$-th time point. Moreover, among all combinations of time points
and channels, $X_{(3,8)}$ has the largest odds ratio (since
$\hat\alpha_3\hat\beta_8$ is the largest among all
$\hat\alpha_i\hat\beta_j$) and would be critical in classifying
alcoholic and control groups. Figure~\ref{eeg_estimates}~(c)
provides the predicted probability of being alcoholism for every
subject (by using the rest 121 subjects) as well as the $95\%$
confidence interval from (\ref{ci.pi}), and
Figure~\ref{eeg_estimates}~(d) gives the kernel density estimates of
$\hat\alpha^TX_i\hat\beta$ for two types of subjects. An obvious
separation of two groups is detected which demonstrates the
usefulness of MV-logistic regression in classification.

The choice of $(p_0,q_0)=(15,15)$ in Step~1 is the same with the
data preprocessing step of Li and others (2010). Under this choice,
we correctly classify 105 of 122 subjects (through leave-one-out
classification procedure) by fitting MV-logistic regression, while
the best result of Li and others (2010) from dimension folding (a
dimension reduction technique that preserves the matrix structure of
covariates) followed by quadratic discriminant analysis gives 97. We
believe the reasons of a better performance for MV-logistic
regression are twofold. First, we adopt a different data
preprocessing technique GLRAM in Step~1, where Li and others (2010)
use a version of (2D)$^2$PCA (Zhang and Zhou, 2005). Hung and others
(2011) show that GLRAM is asymptotically more efficient than
(2D)$^2$PCA in extracting bases, and hence it is reasonable for
GLRAM to produce a better result. Second, standardization in Step~2
makes the EEG Database Data Set more suitable to fit MV-logistic
regression model. Without standardization, MV-logistic regression
cannot produce such a high classification accuracy. This also
reflects that standardization is an important issue before fitting
MV-logistic regression model. We remind the readers again that
standardization of covariates will result in a different MV-logistic
regression model.

For comparison, the $p_0q_0$ extracted covariates in Step~1 are also
fitted with the conventional (penalized) logistic regression of Le
Cessie and Van Houwelingen (1992), where the tuning parameter is
selected in the same way such that the leave-one-out classification
accuracy is maximized. Table~\ref{table_accuracy} provides the
analysis results of both methods under different choices of
$(p_0,q_0)$. One can see that MV-logistic regression uniformly
outperforms conventional logistic regression. Moreover, the
classification accuracy of the conventional approach decays rapidly
as the numbers of $p_0$ and $q_0$ increase, while those of
MV-logistic regression roughly remain constant. As mentioned
previously, MV-logistic regression requires fewer parameters in
model fitting, possesses certain robustness against model violation
and, hence, an efficiency gain is reasonably expected.

\begin{rmk}
As suggested by one referee, we also compare our approach to the
widely used procedure, principal component analysis (PCA) followed
by logistic regression. In particular, PCA is applied to the
vectorized covariates $\vec(X_i^*)$. The leading $r$ principal
components are then used to fit the conventional (penalized)
logistic regression model, where the tuning parameter is also
selected to maximize the leave-one-out classification accuracy (see
Table~\ref{table_pca} for the results). It can be seen this widely
applied approach can not produce classification accuracy higher than
$0.820$, while the best result of MV-logistic regression is $0.861$.
This reveals the limitation of the conventional $\vec(X)$-based
approach which usually produces a large number of parameters.
\end{rmk}

%For comparison, the $p_0q_0$ extracted covariates in Step~1 are also
%used to fit the penalized logistic regression of Le Cessie and Van
%Houwelingen (1992). Table~\ref{table_accuracy} provides the optimal
%leave-one-out classification accuracies of both methods over various
%choices of $(p_0,q_0)$. One can see that MV-logistic regression
%uniformly outperforms conventional logistic regression,
%except for
%the case of $(p_0,q_0)=(10,10)$,
%and attains the maximal accuracy 0.869 at $(p_0,q_0)=(14,14)$,
%wherein 106 out of 122 subjects are correctly classified. Note that
%for $(p_0,q_0)=(10,10)$, as the number of parameters is relatively
%small, conventional logistic regression model has the potential to
%perform well. When $p_0q_0$ excess the available sample size,
%however, conventional approach is not able to extract more
%information, and can not produce classification accuracy higher than
%0.820. Moreover, the classification accuracy of conventional
%approach decays rapidly as the numbers of $p_0$ and $q_0$ increase,
%while those of MV-logistic regression roughly remain constant. This
%reveals that MV-logistic regression is able to extract those extra
%information for larger $(p_0,q_0)$ values. As mentioned previously,
%MV-logistic regression preserves the matrix information of $X$,
%requires less parameters in model fitting, possesses certain
%robustness against model violation and, hence, an efficiency gain is
%reasonably expected.

\section{Extension to Multi-Class Response}
\label{sec6}

We illustrate the extension of MV-logistic regression to the case of
$H$ classes with $H\geq 2$. Let $\tilde Y_i\in\{1,\cdots,H\}$ be the
random variable indicating which class the $i$-th subject belongs
to. We can equivalently code $\tilde Y_i$ as
$Y_i=(Y_{1i},\cdots,Y_{H-1,i})^T$, where $Y_{hi}=1$ indicates the
$i$-th subject belongs to category $h$, $h=1,\cdots,H-1$, and
$\{Y_{hi}=0,h=1,\cdots, H-1\}$ means the $i$-th subject belongs to
category $H$. Consider the model
\begin{eqnarray}
\log\left\{\frac{P(Y_i=h|X)}{P(Y_i=H|X)}\right\}=\gamma_h+\alpha_h^TX\beta_h,~h=1,\cdots,
H-1,
\end{eqnarray}
with $\alpha_h=(1,\alpha_h^{*T})^T$ for the sake of identifiability
as before. The parameter of interest is
$\thetab=(\theta_1^T,\cdots,\theta_{H-1}^T)^T$ with
$\theta_h=(\gamma_h,\alpha_h^{*T},\beta_h^T)^T$. This gives the
covariate-specific probability $P(Y_i=h|X_i)$ to be
$\pi_{hi}=\pi_{h}(\thetab|X_i)=\exp(\gamma_h+\alpha_h^TX_i\beta_h)\{1+\sum_{h=1}^{H-1}\exp(\gamma_h+\alpha_h^TX_i\beta_h)\}^{-1}$
for $h=1,\cdots,H-1$, and
$\pi_{Hi}=\pi_{H}(\thetab|X_i)=\{1+\sum_{h=1}^{H-1}\exp(\gamma_h+\alpha_h^TX_i\beta_h)\}^{-1}$
for $h=H$. Based on the data $\{(Y_i,X_i)\}_{i=1}^n$, the
log-likelihood function of $\thetab$ is derived to be
\begin{eqnarray}
\ell^*(\thetab)=\sum_{i=1}^n\left[\sum_{h=1}^{H-1}Y_{hi}(\gamma_h+\alpha_h^TX_i\beta_h)-
\log\{1+\sum_{h=1}^{H-1}\exp(\gamma_h+\alpha_h^TX_i\beta_h)\}\right].
\end{eqnarray}
Then, $\thetab$ is proposed to be estimated by
$\hat\thetab_{\lambda}=\arg\max_{\thetab}\ell_{\lambda}^*(\thetab)$,
where $\ell_{\lambda}^*(\thetab)=\ell^*(\thetab)-\lambda
J(\thetab)$.

To apply the Newton method to obtain $\hat\thetab_{\lambda}$,  we
need to calculate the gradient vector and Hessian matrix of
$\ell_{\lambda}^*(\thetab)$. Let
$\Xb^*(\thetab)=\text{diag}\{\Xb(\theta_1),\cdots,\Xb(\theta_{H-1})\}$,
where $\Xb(\cdot)$ is defined in the beginning of
Section~\ref{sec3}. Let also $\Yb_h=(Y_{h1},\cdots,Y_{hn})^T$,
$\Pib_h(\thetab)=(\pi_{h1},\cdots,\pi_{hn})^T$ for $h=1,\cdots,H-1$,
$\Yb^*=(\Yb_1^T,\cdots,\Yb_{H-1}^T)^T$, and
$\Pib^*(\thetab)=\{\Pib_1(\thetab)^T,\cdots,\Pib_{H-1}(\thetab)^T\}^T$.
Then, the gradient is derived to be
$\ell^{*(1)}_{\lambda}(\thetab)=\Xb^*(\thetab)^T\{\Yb^*-\Pib^*(\thetab)\}-\lambda
J^{(1)}(\thetab)$,
%\begin{eqnarray}
%\ell^{*(1)}_{\lambda}(\thetab)=\Xb^*(\thetab)^T\{\Yb^*-\Pib^*(\thetab)\}-\lambda
%J^{(1)}(\thetab).\label{grad.multi}
%\end{eqnarray}
and the Hessian matrix (after ignoring the zero expectation term as
in (\ref{hess})) is given by
$H_{\lambda}^*(\thetab)=\Xb^*(\thetab)^T\Vb^*(\thetab)\Xb^*(\thetab)+\lambda
J^{(2)}(\thetab)$,
%\begin{eqnarray}
%H_{\lambda}^*(\thetab)=\Xb^*(\thetab)^T\Vb^*(\thetab)\Xb^*(\thetab)+\lambda
%J^{(2)}(\thetab), \label{hess.multi}
%\end{eqnarray}
where $\Vb^*(\thetab)=[\Vb_{ij}(\thetab)]$,
$\Vb_{hh}=\diag\{\pi_{h1}(1-\pi_{h1}),\cdots,\pi_{hn}(1-\pi_{hn})\}$,
$h=1,\cdots,H-1$, and
$\Vb_{hk}=\Vb_{kh}=\diag(-\pi_{h1}\pi_{k1},\cdots,-\pi_{hn}\pi_{kn})$,
$1\leq h\neq k\leq H-1$. Finally, $\hat\thetab_{\lambda}$ is
obtained by replacing (\ref{grad}) and (\ref{hess}) with
$\ell^{*(1)}_{\lambda}(\thetab)$ and $H_{\lambda}^*(\thetab)$ in the
iteration (\ref{newton}).

By a similar argument of Theorem~\ref{normality}, we can also deduce
that $\sqrt{n}(\hat\thetab_\lambda-\thetab)$ converges weakly to a
normal distribution with mean zero and covariance matrix
$\Sigma^*(\thetab)=\{I^*(\thetab)\}^{-1}$, where
$I^*(\thetab)=E[n^{-1}\Xb^*(\thetab)^T\Vb^*(\thetab)\Xb^*(\thetab)]$.
Moreover, $\Sigma^*(\thetab)$ can be estimated by
$\hat\Sigma^*(\hat\thetab_\lambda)$, where
\begin{eqnarray}
\hat\Sigma^*(\thetab)=\left(\frac{1}{n}H_\lambda^*(\thetab)\right)^{-1}\left(\frac{1}{n}\Xb^*(\thetab)^T\Vb^*(\thetab)\Xb^*(\thetab)\right)
\left(\frac{1}{n}H_\lambda^*(\thetab)\right)^{-1}.
\end{eqnarray}
Inference procedures are the same with what we have already
established in the previous sections.

%\begin{eqnarray}
%\Vb=\left[\begin{array}{cccc}\Vb_1 & \Vb_{12} & \cdots &
%\Vb_{1,H-1}\\ \Vb_{12} & \Vb_{2} & \cdots & \Vb_{2,H-1}\\
%\vdots & \vdots &  & \vdots\\ \Vb_{1,H-1} & \Vb_{2,H-1} & \cdots &
%\Vb_{H-1}\\\end{array}\right].
%\end{eqnarray}

%$\Yb=\left[\begin{array}{c} \Yb_{1}\\ \vdots \\
%\Yb_{H-1}\end{array}\right]$, $\Yb_h=\left[\begin{array}{c} Y_{h1}\\ \vdots \\
%Y_{hn}\end{array}\right]$, $\Pib(\theta)=\left[\begin{array}{c} \Pib_{1}\\ \vdots \\
%\Pib_{H-1}\end{array}\right]$, $\Pib_h=\left[\begin{array}{c} \pi_{h1}\\ \vdots \\
%\pi_{hn}\end{array}\right]$, $h=1,\cdots H-1$,

%$\Xb(\theta_h)=\left[~\1_n,\Xb_0(\beta_h\otimes
%\Cb_p),\Xb_0(\Ib_q\otimes\alpha_h)~\right]$ and

\section{Conclusions}
\label{sec7}

Besides the EEG example, tensor objects are frequently encountered
in many applications, such as mammography images, ultrasound images,
and magnetic resonance imaging (MRI) images, which are examples of
order-two tensors. One can also imagine a subject with $p$
covariates measured on $q$ time points under $t$ different
treatments is naturally stored as an order-three tensor. In this
paper we propose MV-logistic regression model, when the covariates
have a natural matrix structure (order-two tensor). Its performance
is validated through the EEG Database Data Set, where in the best
case of $(p_0,q_0)=(15,15)$ we successfully classify 105 of 122
subjects to the right group. The superiority of MV-logistic
regression comes from the fact that it aims to estimate the best
rank-1 approximation of the true parameter matrix and the parsimony
of parameters used. It is also found in our simulation studies that
MV-logistic regression has certain robustness against the violation
of model specification. Thus, MV-logistic regression model can be
used as a good ``working'' model, especially when the sample size is
relatively small in comparison with the number of covariates.
Although we focus on matrix covariate in this article, the proposed
method can be straightforwardly extended to tensor objects of higher
order.

As discussed in Section~2.2, MV-logistic regression model aims to
approximate the true parameter matrix in the sense of minimum
KL-divergence. Here the discussion of MV-logistic regression model
is only focused on the case of rank-1 approximation due to its
simplicity in implementation and theoretical development, and we
find this simple rank-1 model is suitable for the EEG Database Data
Set. Of course we may encounter situations when rank-1 approximation
is not adequate. It is therefore natural to extend this rank-1
structure to a more general rank-$r$ setting. In particular, for any
positive integer $r\leq p\wedge q$, consider the model
\begin{eqnarray}\label{model.r}
{\rm logit}\{P(Y=1|X)\}=\gamma+\vec(AB^T)^T\vec(X),
\end{eqnarray}
where $A\in \mathbb{R}^{p\times r}$ and $B\in \mathbb{R}^{q\times
r}$ are the row and column parameter matrices, respectively. Note
that model~(\ref{model.r}) can be treated as the conventional
logistic regression model~(\ref{model.full}) with the constraint
$\xi=AB^T$, and will reduce to our MV-logistic regression
model~(\ref{model}) when $r=1$. For the general rank-$r$ setting,
however, some issues need to be further studied, such as the problem
of identifiability of parameters in $(A,B)$ and the corresponding
asymptotic properties. Moreover, the algorithm developed in this
paper can not be directly applied, and also need to be revised to
adapt to model~(\ref{model.r}). We believe model~(\ref{model.r}) has
its wide applicability in practice, and is of great interest to
investigate in a future study.

\clearpage

\begin{center}
{REFERENCES}
\end{center}

\begin{description}

\item
Bickel, P. J. and Doksum, K. A. (2001). \textit{Mathematical
Statistics: Basic Ideas and Selected Topics Vol. I}, 2nd edition.
New Jersey: Prentice-Hall.

\item
Green, P. J. (1984). Iteratively reweighted least squares for
maximum likelihood estimation, and some robust and resistant
alternatives (with discussion). {\em Journal of the Royal
Statistical Society: Series B} \textbf{46}, 149-192.

\item
Hastie, T., Tibshirani, R., and Friedman, J. (2009).\textit{The
Elements of Statistical Learning: Data Mining, Inference, and
Prediction}, 2nd edition. Berlin: Springer.

\item Hung, H., Wu, P. S., Tu, I. P.,
and Huang, S. Y. (2011). On multilinear principal component analysis
of order-two tensors, manuscript. arXiv:1104.5281v1.

\item
Le Cessie, S. and Van Houwelingen, J. C. (1992). Ridge estimators in
logistic regression. {\em Journal of the Royal Statistical Society:
Series C} \textbf{41}, 191-201.

\item
Lee, A. and Silvapulle, M. (1988). Ridge estimation in logistic
regression. \textit{Communications in Statistics, Simulation and
Computation} \textbf{17}, 1231-1257.

\item
Li, B., Kim, M. K., and Altman, N. (2010). On dimension folding of
matrix- or array-valued statistical objects. {\em The Annals of
Statistics} \textbf{38}, 1094-1121.

\item
Lu, W. S., Pei, S. C. and Wang, P. H. (1997). Weighted low-rank
approximation of general complex matrices and its application in the
design of 2-D digital filters. \textit{IEEE Transactions on Circuits
and Systems-I} \textbf{44}, 650-655.

\item
Manton, J. H., Mahony, R., and Hua, Y. (2003). The geometry of
weighted low-rank approximations.  {\it IEEE Transactions on Signal
Processing} \textbf{51}, 500-514.

\item
MuCulloch, C. E., Searle, S. R., and Neuhaus, J. M. (2008).
\textit{Generalized, Linear, and Mixed Models}. Wiley: New York.

\item
Park, M. Y. and Hastie, T. (2008). Penalized logistic regression for
detecting gene interactions. \textit{Biostatistics} \textbf{9},
30-50.

\item
Ye, J. (2005). Generalized low rank approximations of matrices. {\it
Machine Learning} \textbf{61}, 167-191.

\item
Zhang, D. and Zhou, Z. H. (2005). ${\rm (2D)}^2$PCA: Two-directional
two-dimensional PCA for efficient face representation and
recognition. {\it Neurocomputing} \textbf{69}, 224-231.

\item
Zhu, J. and Hastie, T. (2004). Classification of gene microarrays by
penalized logistic regression. \textit{Biostatistics} \textbf{5},
427-443.
\end{description}

\clearpage

\begin{table}[t!]
\caption{Averages of $\hat\theta_{\lambda}$ (Mean), averages of
diagonal elements of $\hat\Sigma(\hat\theta_{\lambda})$ (SE), and
standard deviations of $\hat\theta_{\lambda}$ (SD) under
model~(\ref{model}) with $(p,q)=(12,10)$. The last row gives
averages (standard deviations) of similarities (SIM) between
$(\gamma,\beta^T\otimes\alpha^T)^T$ and
$(\hat\gamma,\hat\beta^T\otimes\hat\alpha^T)^T$ \label{sim1}}
{\tabcolsep=4.25pt
\begin{center}
\begin{tabular}{ccccccccc}
\hline
& & \multicolumn{3}{c}{$n = 150$} & \multicolumn{3}{c}{$n = 300$} \\
\cmidrule(l){3-5}\cmidrule(l){6-8}
& True & Mean & SD & SE& Mean & SD & SE\\
\hline
$\gamma  $  & ~1.000 & ~1.095 & 0.447 & 0.460 & ~1.055 & 0.289 & 0.287 \\
\hline
$\alpha^*$  & ~0.500 & ~0.549 & 0.165 & 0.157 & ~0.525 & 0.098 & 0.096 \\
          & -0.500 & -0.543 & 0.169 & 0.156 & -0.524 & 0.103 & 0.096 \\
          & -0.500 & -0.558 & 0.162 & 0.158 & -0.525 & 0.100 & 0.095 \\
          & -0.500 & -0.557 & 0.170 & 0.157 & -0.528 & 0.095 & 0.095 \\
          & -0.500 & -0.560 & 0.168 & 0.157 & -0.524 & 0.101 & 0.096 \\
          & -0.500 & -0.550 & 0.165 & 0.156 & -0.524 & 0.094 & 0.095 \\
          & -0.500 & -0.557 & 0.168 & 0.157 & -0.523 & 0.096 & 0.095 \\
          & -0.500 & -0.546 & 0.161 & 0.157 & -0.525 & 0.098 & 0.096 \\
          & -0.500 & -0.557 & 0.164 & 0.157 & -0.527 & 0.097 & 0.096 \\
          & -0.500 & -0.555 & 0.163 & 0.157 & -0.521 & 0.097 & 0.095 \\
          & -0.500 & -0.554 & 0.159 & 0.156 & -0.526 & 0.102 & 0.095 \\
\hline
$\beta   $  & ~1.000 & ~0.972 & 0.226 & 0.237 & ~0.995 & 0.174 & 0.177 \\
          & ~0.500 & ~0.485 & 0.206 & 0.207 & ~0.506 & 0.141 & 0.143 \\
          & ~1.000 & ~0.962 & 0.220 & 0.238 & ~0.999 & 0.170 & 0.178 \\
          & -1.000 & -0.970 & 0.222 & 0.238 & -1.008 & 0.171 & 0.179 \\
          & -1.000 & -0.974 & 0.219 & 0.238 & -0.999 & 0.172 & 0.178 \\
          & -1.000 & -0.969 & 0.222 & 0.238 & -0.996 & 0.172 & 0.177 \\
          & -1.000 & -0.962 & 0.215 & 0.238 & -1.008 & 0.169 & 0.178 \\
          & -1.000 & -0.968 & 0.223 & 0.238 & -0.996 & 0.169 & 0.178 \\
          & -1.000 & -0.966 & 0.220 & 0.238 & -1.012 & 0.171 & 0.179 \\
          & -1.000 & -0.959 & 0.216 & 0.237 & -1.000 & 0.172 & 0.178 \\
\hline
     SIM     &  & ~0.950 & (0.021) &  & ~0.981 & (0.007) &  \\
\hline
\end{tabular}
\end{center}
}
\end{table}

\begin{table}[t!]
\caption{Averages of classification accuracies (CA) of
MV-logistic/Logistic regression and winning proportion (WP) of
MV-logistic regression for different values of $\sigma$, $p$, $q$
and $n=150$. $\rho_\sigma$ is the average of the explained
proportion of the best rank-1 approximation. \label{sim3}.}
{\tabcolsep=4.25pt
\begin{center}
\begin{tabular}{ccccccc}
\hline
$(p,q)$ &     \multicolumn{2}{c}{$(12,10)$}   &  \multicolumn{2}{c}{$(20,15)$}  &   \multicolumn{2}{c}{$(20,30)$}   \\
\cmidrule(r){1-1}
\cmidrule(lr){2-3}\cmidrule(rl){4-5}\cmidrule(l){6-7}
$\sigma$    &   $\rho_\sigma$ &   CA~(WP)    &   $\rho_\sigma$   &  CA~(WP)    &   $\rho_\sigma$ &   CA~(WP)    \\
\cmidrule(r){1-1}
\cmidrule(lr){2-3}\cmidrule(rl){4-5}\cmidrule(l){6-7}
0   &   100\%    &   0.867/0.739~(1.00) &   100\%    &   0.866/0.670~(1.00) &      100\%    &   0.839/0.622~(1.00) \\
0.1 &   69\% &   0.856/0.741~(1.00) &   63\% &   0.848/0.668~(1.00) &   58\% &   0.821/0.622~(1.00) \\
0.3 &   44\% &   0.794/0.745~(0.87) &   36\% &   0.765/0.669~(0.94) &   32\% &   0.720/0.623~(0.94) \\
0.5 &   33\% &   0.727/0.751~(0.37) &   26\% &   0.677/0.673~(0.60) &   22\% &   0.631/0.621~(0.61) \\
\hline
\end{tabular}
\end{center} }
\end{table}

\begin{table}[t!]
\caption{The leave-one-out classification accuracy of
MV-logistic/Logistic regression for the EEG Database Data Set under
different combinations of $(p_0,q_0)$ \label{table_accuracy}.}
{\tabcolsep=4.25pt
\begin{center}
\begin{tabular}{ccccccc}
 \hline
    & \multicolumn{3}{c}{$q_0$}\\
    \cmidrule(lr){2-4}
  $p_0$  & 15 & 20 & 30 \\
\hline
15 &   \textbf{0.861}/0.803 &   0.836/0.795 &   0.844/0.787      \\
30 &   0.844/0.779 &   0.828/0.779 &   0.828/0.754      \\
60 &   0.844/0.737 &   0.853/0.713 &   0.828/0.713      \\
\hline
 \end{tabular}
 \end{center}
}
\end{table}

\begin{table}[t!]
\caption{The leave-one-out classification accuracy of PCA followed
by conventional logistic regression for the EEG Database Data Set
under different choices of $r$ \label{table_pca}.}
{\tabcolsep=4.25pt
\begin{center}
\begin{tabular}{ccccccccccc}
\hline
$r$ & 1 & 2 & 3 & 4 & 5 & 6 & 7 & 8 & 9 & 10 \\
Accuracy & 0.631  & 0.648 & 0.730 & \textbf{0.820} & \textbf{0.820} & 0.803 & 0.787 & 0.779 & 0.787 & 0.795 \\
\hline
$r$ & 20 & 30 & 40 & 50 & 60 & 70 & 80 & 90 & 100 & 120 \\
Accuracy & 0.746 & 0.779 & 0.795 & 0.754 & 0.738 & 0.746 & 0.713 & 0.705 & 0.631 & 0.492 \\
\hline
\end{tabular}
\end{center} }
\end{table}

\end{document}